\documentclass[%
 reprint,
 amsmath,amssymb,
 aps,onecolumn,
]{revtex4-2}

\usepackage{amsmath}
\usepackage{graphicx}% Include figure files
\usepackage{dcolumn}% Align table columns on decimal point
\usepackage{bm}% bold math
\begin{document}

\preprint{APS/123-QED}

\title{Superposition of system response in modulated turbulent plane Couette flow}

\author{M. Wasy Akhtar}
\affiliation{Element Digital Engineering, \\14805 Yorktown Plaza Drive, Houston, TX, 77040, USA}

\author{Rodolfo Ostilla-M\'onico}
\email{rodolfo.ostilla@uca.es}
\affiliation{Dpto.~Ing. Mec\'anica y Dise\~no Industrial, Escuela Superior de Ingenier\'ia, Universidad de C\'adiz, Av.~de la Universidad de C\'adiz 10, 11519 Puerto Real, Espa\~na}

\date{\today}% It is always \today, today,

\begin{abstract}
Traditionally, the response of a turbulent flow to modulated perturbations is expected to be complex. We conduct direct numerical simulations of turbulent Plane Couette flow, the shear flow between two differentially moving plates, to reveal intriguing behaviour in response to a modulation in the velocity of one of the plates: the response of the flow to modulations can be calculated from a superposition of the response to each Fourier mode of the modulation. We fix the shear Reynolds number as $Re_s=3\times10^4$, and measure the propagation of a modulated forcing in the flow close to the resonant frequency of the system. We find that the amplitude and phase response of the flow can be largely captured using models based on laminar flow, regardless of the waveform used to force the flow. Furthermore, we find that linear superposition can effectively describe the flow's response to these perturbations as long as the modulation is smaller than the base flow. Our findings unveil the persistence of linear superposition in a turbulent flow under specific conditions, even when perturbations occur at time-scales closely aligned with the system's characteristic time-scales.
\end{abstract}


\maketitle

\section{Introduction}

%Non-linear systems pervade both natural and societal phenomena, and span diverse scales from the cosmic to the microscopic. Despite their prevalence, understanding and predicting the behavior of such systems remains a formidable challenge. The inherent complexity of non-linear dynamics often leads to unexpected consequences, where seemingly minor disturbances can trigger significant and unpredictable outcomes. As a consequence of this, the search for regimes where non-linear systems behave in a less chaotic way and can be modelled and controlled is a very active interdisciplinary field ranging from ecology and neurology \cite{bairey2016high, vlachas2022multiscale, battiston2021physics} to laser physics and machine learning \cite{battiston2020networks, loirette2021linearized}. 

Hydrodynamical turbulence is one of the paradigmatic examples used to explore the physics of non-linear systems. The Navier-Stokes equations describe the behavior of fluids, and include non-linear and non-local terms which are usually responsible for creating the broad-band nature of turbulence \cite{waleffe1992nature}. However, under certain circumstances fluid flows or flow regions may respond in a linear or quasi-linear manner, such as rotating flows or those near the laminar-turbulent transition \cite{oishi2023generalized, tobias2017three}. While instances where turbulence behaves (quasi-)linearly as a whole are rare, they offer invaluable insights into the nature and understanding of turbulence which can be used to develop simpler models \cite{jimenez2013linear,mckeon2017engine, skouloudis2021scaling}.

In particular, shear flows present an interesting case as the maintenance of turbulence in the near-wall region has been related to linear processes \cite{kim2000linear}. By focusing on the large-scales in a turbulent flow, several authors have been able to analyze how the different length- and time-scales in a shear flow interact producing perturbation growth and sustaining turbulence \cite{bamieh2001energy,jovanovic2005componentwise,mckeon2010critical,farrell2014statistical,zare2017colour}, lending support to the idea that most of the physics behind the ``engine'' of wall-turbulence can be captured through adequately linearized Navier-Stokes equations \cite{mckeon2017engine}.

Here, we approach the problem of the linear behaviour of shear flows from another perspective. One effective method for probing the extent of non-linearity of a system is by subjecting it to modulated external forcing and measuring the response. Typically, one does not expect a fully turbulent flow to generally respond to a modulated perturbation in a manner which can be captured through a reduced or linearised models due to the non-linear and broad-band nature of turbulence, i.e.~ in general, a superposition of periodic oscillations with different frequencies cannot explain the fluid system dynamics \cite{Landau:1944ibh,Eckmann1985ErgodicTO}. For example, in homogeneous isotropic turbulence (HIT), as the modulation frequency approaches internal flow time scales, the turbulence field adjusts to the applied modulation, and a dominant scale correlated to the forcing frequency arises which gives rise to couplings between the forcing and existing turbulent structures, resulting in phenomena like amplified energy injection and dissipation \cite{von2003response67,von2003response68,bos2007small,kuczaj2008turbulence,cekli2010resonant,cekli2015stirring}. In thermal convection, substantially enhanced heat transfer can be achieved by perturbing the boundary conditions of the system at frequencies close to the characteristic frequency of the large-scale flow structures \cite{jin2008experimental}.

Therefore, predicting the response of a turbulent flow to a modulated perturbation appears in first instance as a formidable challenge. However, as the general dynamics of shear flows can be captured using simpler and linearized models \cite{mckeon2017engine}, shear flows may present different responses to modulation than those mentioned above. Indeed, our earlier work \cite{akhtar2022effect} showed that the response of Plane Couette flow (PCF), i.e.~the flow between two differentially moving plates, to a small sinusoidal modulation in the driving shows a very similar behaviour as the analytic solution in the absence of turbulence, and that unlike HIT, the large-scale structures in the flow remained relatively unaffected by the modulation. In this manuscript, we explore these ideas further by subjecting PCF to arbitrary modulations with large amplitude. Remarkably, we show that linear superposition can effectively describe the response of the flow to any perturbation: once the response to a sinusoidal perturbation is known, the response of the flow to an arbitrary perturbation can be be constructed through superposition of the individual Fourier components, provided the perturbations do not exceed the magnitude of the underlying forcing. Despite the system's inherent non-linearity, linear superposition persists in wall-bounded turbulent flows under certain conditions even when forcing occurs at time-scales close to the system's characteristic time-scales.

\section{Numerical method}

To simulate PCF, we use a three-dimensional Cartesian domain which is periodic in the $x$ (streamwise) and $z$ (spanwise) directions with periodicity lengths $L_{x}$ and $L_{z}$ respectively, and is bounded in the $y$ direction by two plates separated a distance $d$. The velocity at the top and bottom plates are equal and opposite $\pm (U/2) \textbf{e}_x$. In addition, an arbitrary zero-mean modulation is superimposed onto the bottom plate's base velocity.

The incompressible non-dimensional Navier-Stokes equations which govern this problem are:

\begin{equation}
 \frac{\partial \textbf{u}}{\partial t} + \textbf{u}\cdot \nabla \textbf{u} = -\nabla p + Re_s^{-1} \nabla^2 \textbf{u},
  \label{eq:ns1}
\end{equation}
 
\noindent which alongside the incompressibility condition defines the flow field,
 
\begin{equation}
 \nabla \cdot \textbf{u} = 0.
 \label{eq:ns2}
\end{equation}

\noindent Here $\textbf{u}$ is the non-dimensional velocity, $p$ the pressure and $t$ is time. The equations \ref{eq:ns1}-\ref{eq:ns2} are non-dimensionalized using the gap-width $d$ and the plate velocity $U$, resulting in a non-dimensional control parameter: the shear Reynolds number, $Re_s=Ud/\nu$, which can be seen as the non-dimensional strength of the shear forcing. In this manuscript, the Reynolds number $Re_s$ is fixed to $3 \times 10^4$, resulting in a frictional Reynolds number $Re_\tau=u_\tau d/(2\nu)\approx 400$. Here, $u_\tau$ is the shear velocity defined as $u_\tau = \sqrt{\tau_w/\rho}$, where $\tau_w$ is the average wall shear stress and $\rho$ is the fluid density. For PCF, this value of $Re_\tau$ results in a flow with well-developed turbulence \cite{avsarkisov2014turbulent,pirozzoli2014turbulence}. 

As mentioned above, a modulation is superimposed to the boundary conditions such that the bottom plate has a (non-dimensional) velocity of $\textbf{u}(y=-1/2)=[1/2+f(t)]\textbf{e}_x$, where $f(t)$ is an periodic zero-mean function with maximum non-dimensional amplitude $\alpha$ and non-dimensional period $T$. The period can also be expressed in wall-units by using the turbulent time-scale $d/u_\tau$, such that $T_\tau = T U/u_\tau$. This time-scale is closer to that of the flow fluctuations, and hence a better choice for expressing the modulation \cite{akhtar2022effect}. We also note that due to the high Reynolds numbers of this study, this choice of time-scale is more appropriate than the viscous scale $d^2/\nu$ which produces the Womersley number often-used to non-dimensionalize pulsatile flows at low Reynolds numbers \cite{ling1972nonlinear,ku1983pulsatile}. In this manuscript, both the modulation's non-dimensional amplitude $\alpha$ and shape (e.g.~sawtooth, square wave,...) are systematically varied to investigate their impact on the flow response. Unless otherwise stated, the period of the modulation is chosen as $T_\tau=0.8$, close to the characteristic time-scale of the system. A few simulations are conducted with $T_\tau=0.4$ to demonstrate the independence of the results on the modulation frequency. Periodic aspect ratios of $L_x/d=2\pi$ and $L_z/d=\pi$ are used in this study. As this is generally considered a small aspect ratio for PCF \cite{avsarkisov2014turbulent}, additional simulations are conducted at $L_x/d=4\pi$ and $L_z/d=2\pi$ to ensure box-size independence. These are reported in the Appendix. 

Equations \ref{eq:ns1}-\ref{eq:ns2} are advanced in time using a second-order energy-conserving finite difference code known as AFiD \cite{van2015pencil}, which has been used previously for such shear flows \cite{ostilla2016near,akhtar2022effect}. The spatial resolution is set as $512\times384\times512$, which was validated in previous studies at this Reynolds number \cite{akhtar2022effect}. Simulations are started from zero initial conditions, and after an initial time of at least five cycles (to exclude any initial transients), statistics are collected over $50$ modulation periods. Key metrics such as the flow velocities and the energy dissipation rate are recorded for analysis.

%% ------ %%

\section{Results} 

\begin{figure}
  \includegraphics[width=0.32\textwidth]{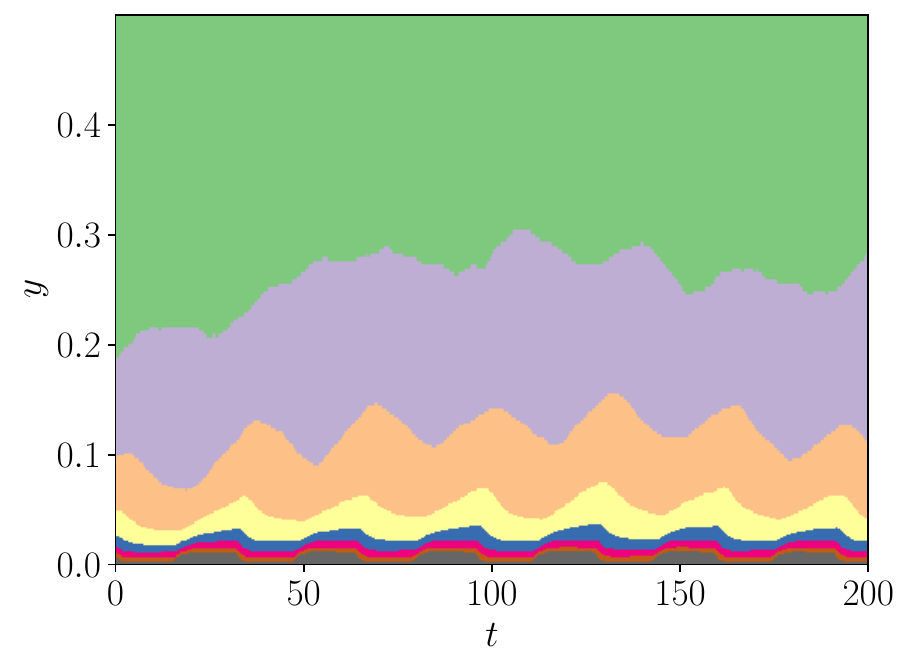}
  \includegraphics[width=0.32\textwidth]{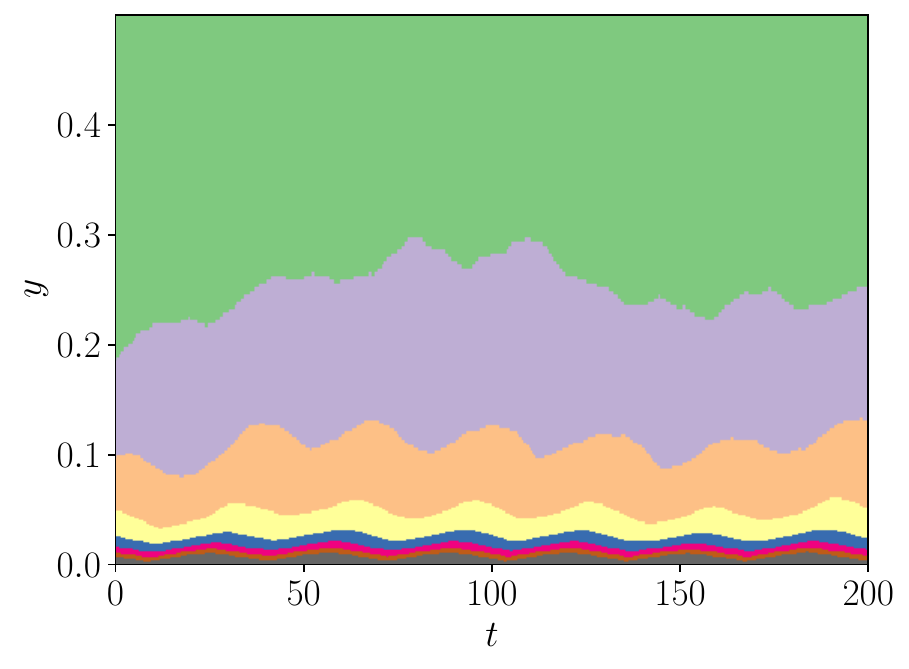}
  \includegraphics[width=0.32\textwidth]{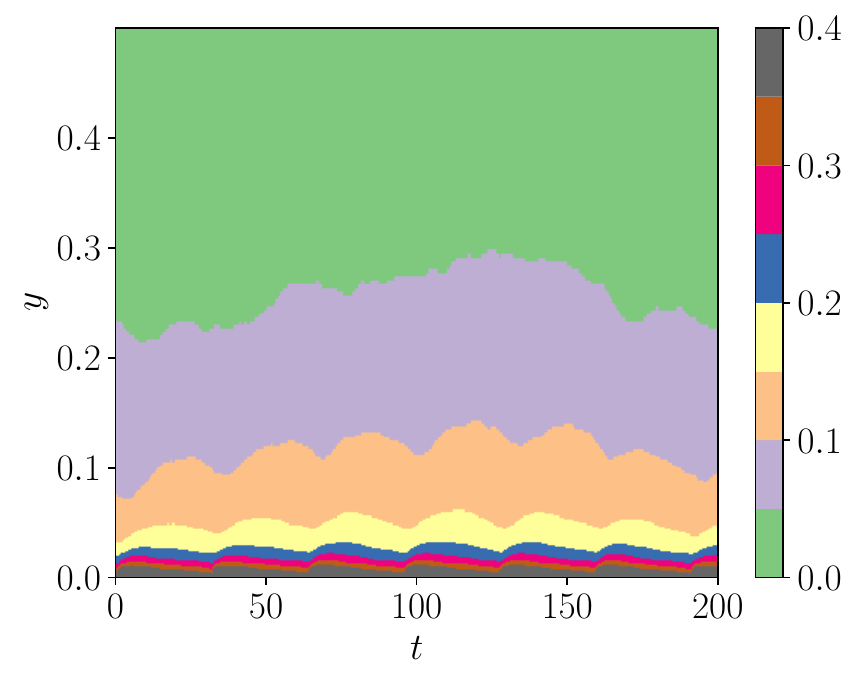}
  \includegraphics[width=0.32\textwidth]{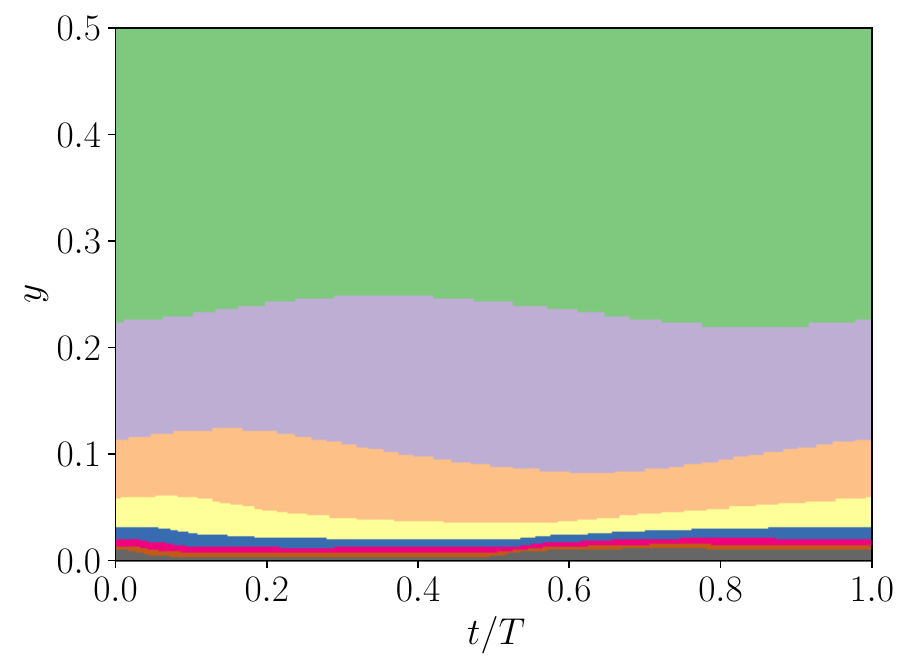}
  \includegraphics[width=0.32\textwidth]{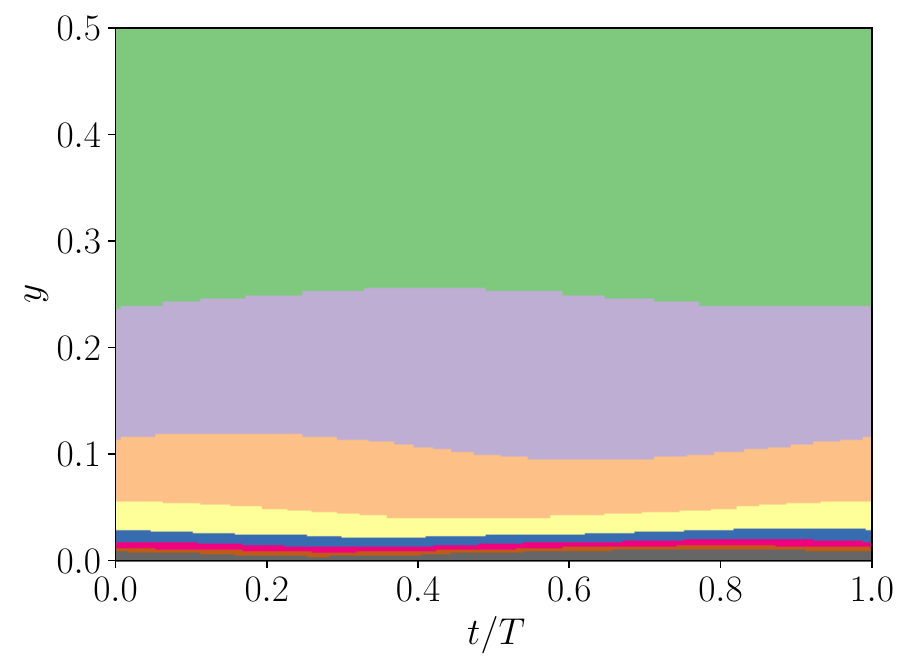}
  \includegraphics[width=0.32\textwidth]{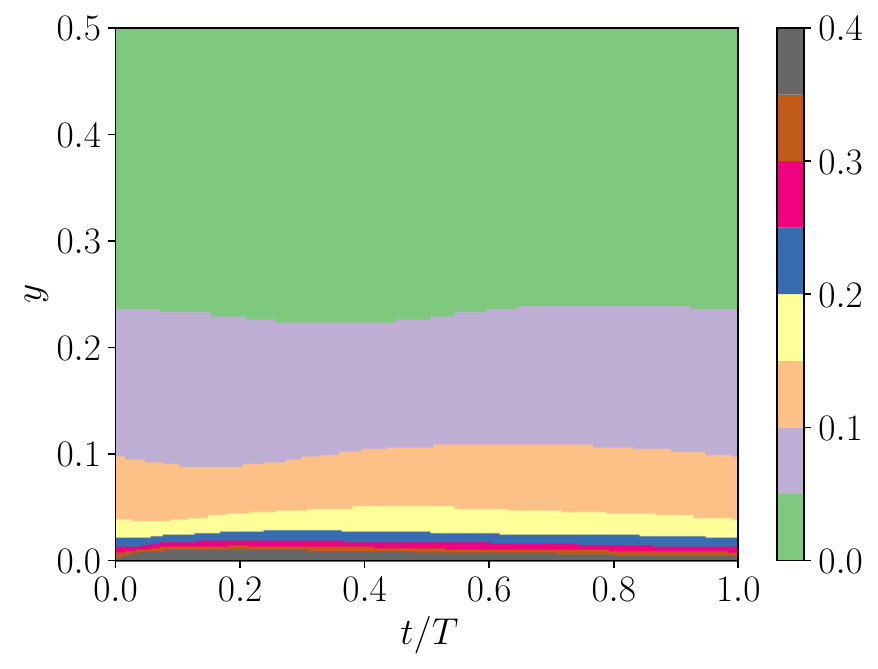}
   \caption{Top row: Spanwise and streamwise averaged streamwise velocity $\langle u_x \rangle_{x,z}$ up to the mid-gap for a square (left), triangle (middle), and sawtooth (right) modulation. Bottom row: phase averaged streamwise velocity $\bar{u}_x$ for the same modulations.}
   \label{fig:velocity_full}
 \end{figure}

First, we fix $\alpha=0.1$ and use three different waveforms for the modulation: square, sawtooth and triangular wave. Despite the ``spikiness'' of the forcing, which is considered a critical parameter to obtain non-linear responses in analogous wall-bounded systems such as thermal convection \cite{jin2008experimental}, we observe no discernible impact on the energy dissipation within averaging errors. This matches and extends previous results which employed only sinusoidal modulation \cite{akhtar2022effect}.

To further showcase the effects of the modulation, we examine the velocity response for different modulations in the top row of figure \ref{fig:velocity_full}, which shows a space-time visualization of the instantaneous stream- and span-wise averaged streamwise velocity, i.e.~$\langle u_x \rangle_{x,z}$, where the operator $\langle .. \rangle_{i}$ denotes averaging with respect to the variable $i$. For all waveforms, close to the wall the velocity follows the modulation imposed on the wall. As the modulation traverses through the flow, the higher frequency components of the forcing gradually dissipate leading to velocity fluctuations with no distinct memory of the modulating waveform. 

To distinguish the effects of turbulence from the effects of modulation, we conduct a phase average of $\langle u_x \rangle_{x,z}$ over several time periods. This results in a periodic velocity field denoted as $\bar{u}_x$, which has a temporal domain of $0\leq t/T < 1$. The bottom row of figure \ref{fig:velocity_full} shows a spacetime visualization of $\bar{u}_x$ for the three waveforms shown in the top row. With the effects of turbulent fluctuations averaged out, the figure clearly shows that as the distance to the wall increases the fluctuations become smoother with a sinusoidal waveform close to the first harmonic.

To analyze this quantitatively, we follow Ref.~\cite{akhtar2022effect} and decompose $\bar{u}_x(y,t/T)$ into its time-averaged component $\bar{U}_{x}(y)$, and a fluctuation $\bar{u}_x^\prime(y,t/T)$, such that:
\begin{equation}
    \bar{u}_x(y,t/T)=\bar{U}_{x}(y)+\bar{u}_x^\prime(y,t/T), \mathrm{ ~ with ~ } \bar{U}_x(y) = \frac{1}{T} \int_0^T \bar{u}_x(y,t/T) dt 
\end{equation}

\noindent Consistent with previous observations for sinusoidal perturbations of similar and smaller amplitude \cite{akhtar2022effect}, we obtain that $\bar{U}_x$ remains unchanged compared to the unmodulated case across all waveforms. This shows that both sinusoidal perturbations with a single period, and more complex waveforms which force a combination of many Fourier modes simultaneously, do not interact with the broadband structures of the flow. Instead, their effect averages out. This lack of coupling is distinct from what one could naively expect from what is observed in other turbulent flows such as homogeneous isotropic turbulence or thermal convection \cite{cekli2010resonant,jin2008experimental}.

We can further decompose $\bar{u}_x^\prime$ into Fourier modes:
\begin{equation}
    \bar{u}_x^\prime(y,t/T) = \sum_{n=1}^\infty A_n(y) \sin(nt/T + \phi_n(y)),
\end{equation}
\noindent Both the amplitude $A_n$ and the phase $\phi_n$ are functions of wall-distance $y$. This allows us to examine the behavior of specific Fourier modes in the flow. In particular, $A_1(y)$ and $\phi_1(y)$, represent how the fundamental mode propagates into the flow from the wall. 

\begin{figure}
  \includegraphics[width=0.45\textwidth]{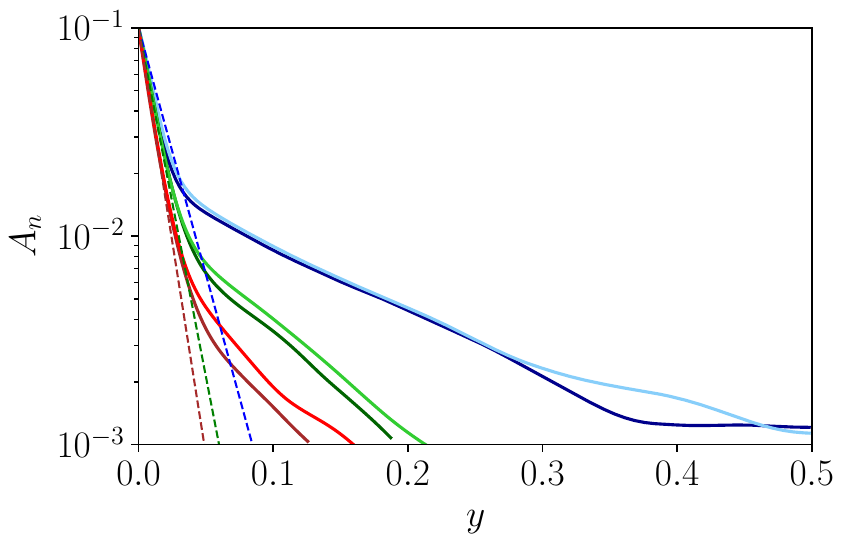}
  \includegraphics[width=0.45\textwidth]{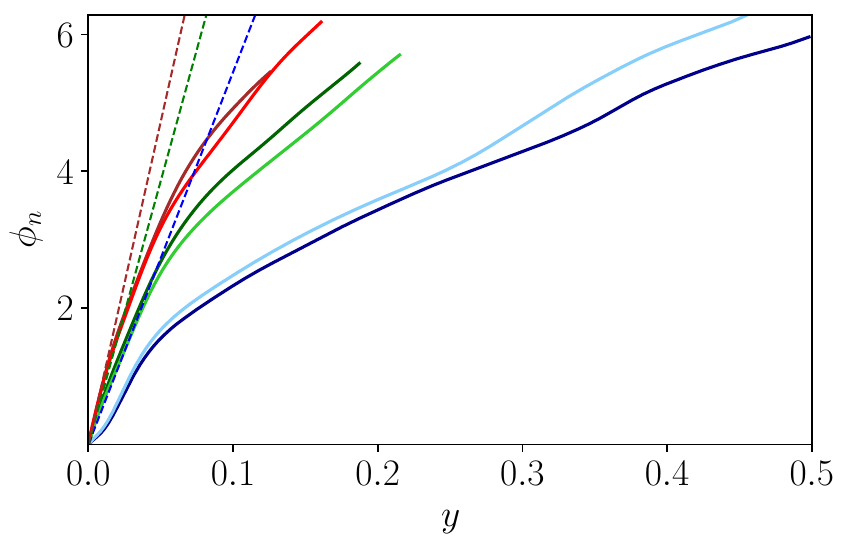}
   \caption{Amplitude (left) and phase (right) response of the first three modes for the 1-2-3 simulation (light lines) and equivalent simulations with just one amplitude. Symbols: blue, fundamental model; green, second harmonic; red, third harmonic. Solid dashed lines indicate the respective Stokes solution. The phase $\phi_n$ is truncated when $A_n<10^{-3}$ as it becomes dominated by noise.}
   \label{fig:a123_phi123_wo77}
 \end{figure}

These variables are shown in Figure \ref{fig:a123_phi123_wo77} for the three waveforms considered, a sinusoidal forcing at the same frequency, and the ``1-2-3 waveform'': a modulation consisting of a linear combination of three sine waves at the fundamental frequency, the second and the third harmonic. Remarkably, we find that the amplitude $A_1$ and phase response $\phi_1$ of the fundamental mode are consistent across all waveforms. Consistent with Ref.~\cite{akhtar2022effect}, they exhibit two distinct regions: a boundary-layer region roughly adhering to laminar theoretical solutions and a bulk-response demonstrating behavior akin to the laminar problem albeit with increased turbulent viscosity. Deviations only start to appear once $A_1\simeq10^{-3}$, which was previously considered as the precision limit of the simulations \cite{akhtar2022effect}.

\begin{figure}
  \includegraphics[width=0.45\textwidth]{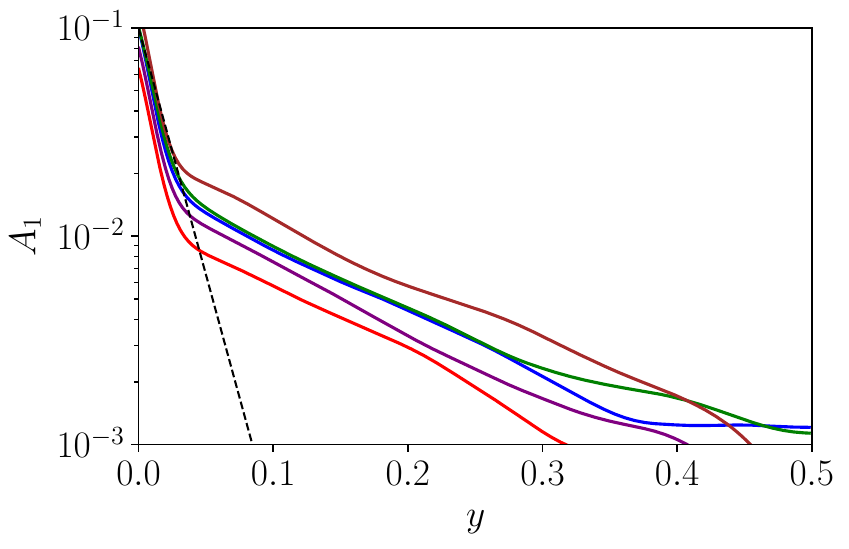}
  \includegraphics[width=0.45\textwidth]{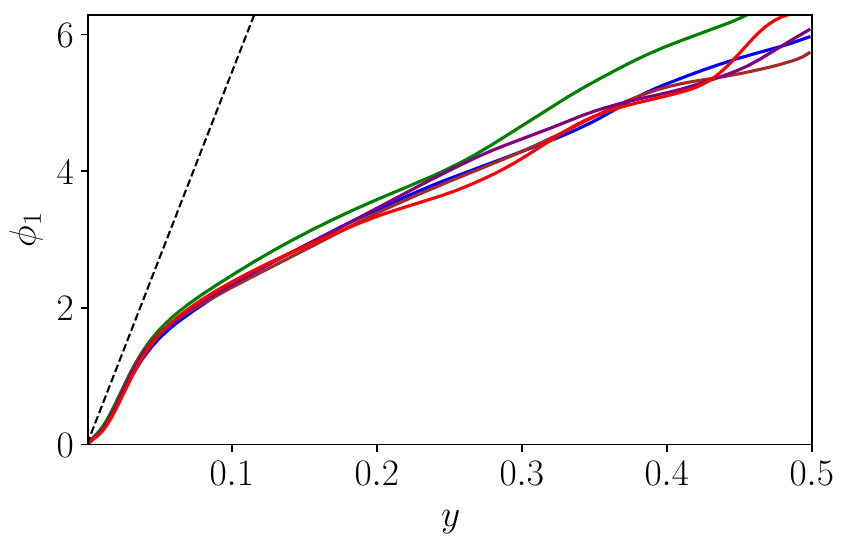}
   \caption{Amplitude (left) and phase (right) response of the fundamental mode for all waveforms at $T_\tau=0.8$. Symbols: blue, sinusoidal modulation at the fundamental frequency;  brown, square wave; purple, triangle wave; red, saw-tooth wave; green, 1-2-3 simulation. Solid dashed lines indicate the theoretical laminar (Stokes) solution.}
   \label{fig:a1_phi1_wo77_all}
 \end{figure}

This result suggests two things: first, as previously reported, the response of the flow to sinusoidal modulations is captured by the physics of the laminar problem \cite{akhtar2022effect}. Secondly, that the response to more complex waveforms can be understood from the responses to individual sinusoids. That is, the response of this non-linear system to modulated forcing entails the superposition of individual responses to the sinusoidal components of the forcing. This also clarifies why the wave-form of the perturbations become smoothed out as the distance from the wall increases: as the perturbation moves into the flow through the boundary layer, higher frequencies are damped out more rapidly than the smaller ones \cite{akhtar2022effect}. This means that rather than coupling to the existing large-scales and ``kicking'' them in a way which enhances dissipation, as was seen in Refs.~\cite{cekli2010resonant,jin2008experimental}, they are mainly dissipated in the viscous regions of the boundary layer.

\begin{figure}
  \includegraphics[width=0.45\textwidth]{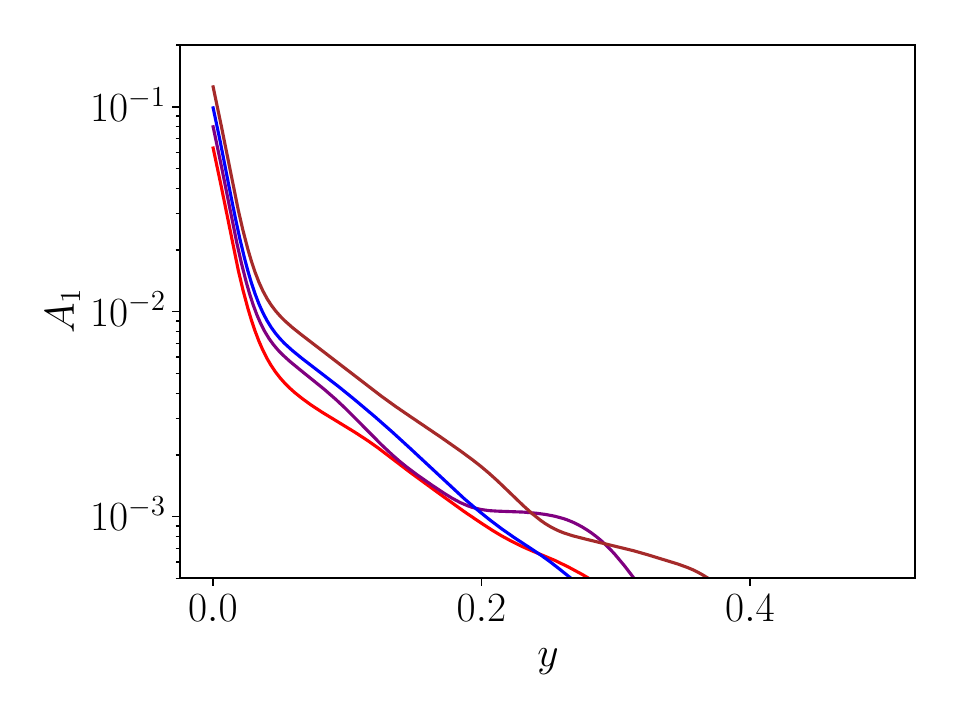}
  \includegraphics[width=0.45\textwidth]{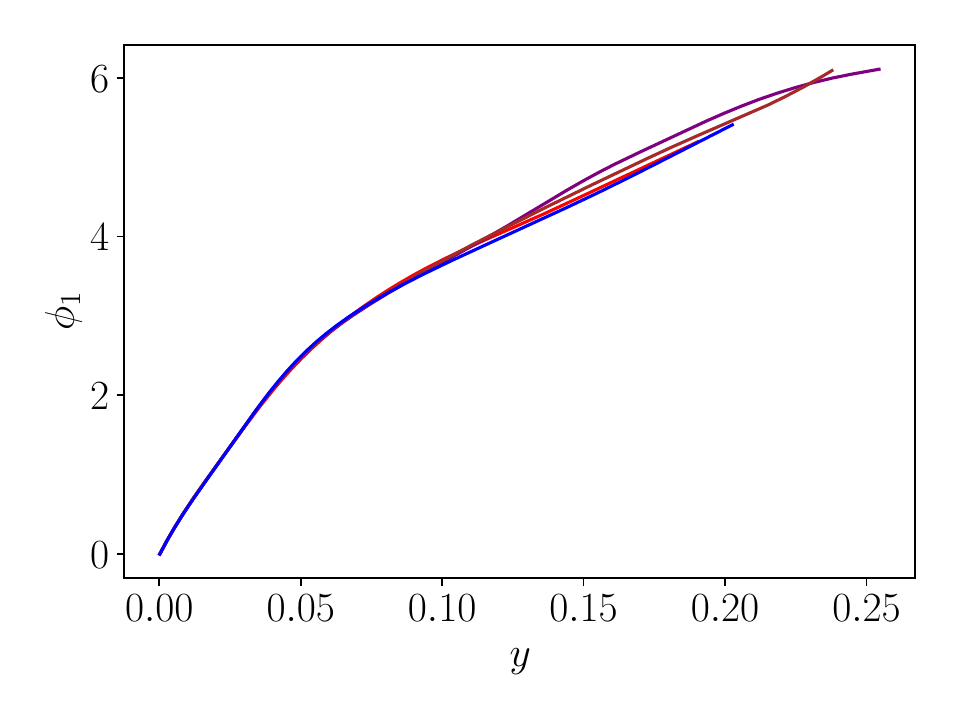}
   \caption{Amplitude (left) and phase (right) response of the fundamental mode for all waveforms at $T_\tau=0.4$. Symbols: blue, sinusoidal modulation at the fundamental frequency;  brown, square wave; purple, triangle wave; red, saw-tooth wave. The phase $\phi_1$ is truncated when $A_1<10^{-3}$ as it becomes dominated by noise.}
   \label{fig:a1_phi1_wo114_all}
 \end{figure}

To ensure that this is phenomena is frequency-independent, we conduct three more simulations at $T_\tau=0.4$, that is, for modulations with twice the frequency as the previous case. The results are shown in Figure \ref{fig:a1_phi1_wo114_all}. First, we notice that the perturbations decay more rapidly, as could be expected from earlier work \cite{akhtar2022effect}. More importantly, as long as the amplitude of the base frequency remains large enough to not be diluted by numerical noise ($A_1\geq10^{-3}$), the simulations confirm what is shown in Figure \ref{fig:a1_phi1_wo77_all}: that the base frequency of the modulation behaves independent of higher harmonics.

To further investigate this characteristic response, we can check if two properties can be satisfied. First, we verify if the response ($A$, $\phi$, ...) to a modulation $f(t)$ which can be expressed as a superposition of two modulations, i.e.~$f(t)=f_a(t)+f_b(t)$, is equal to the superposition of the response of the flow to the two separate modulations. This is done by comparing a simulation where the plate velocity is modulated using a linear combination of three sinusoids of equal amplitude $\alpha=0.1$ and periods $T_{\tau,0}=0.8$, $T_{\tau,0}/2$, and $T_{\tau,0}/3$ (earlier denoted as the ``1-2-3 waveform'') against results from three separate simulations where the plate velocity is modulated with a single sinusoid at each corresponding period $T_{\tau,0}$, $T_{\tau,0}/2$ and $T_{\tau,0}/3$.

The results for the amplitude and phase are shown in Figure \ref{fig:a123_phi123_wo77}. Remarkably, we observe very good agreement in the amplitudes up to $A\sim5\times10^{-3}$, corresponding to half of a percent accuracy, which is the accuracy of our simulations. Similarly, the phase also demonstrates a high degree of agreement between the 1-2-3 waveform and the simulations which use a single sinusoid. This confirms that the response of the system to a combined forcing can be represented as the superposition of the individual responses. 

\begin{figure}
 \includegraphics[width=0.45\textwidth]{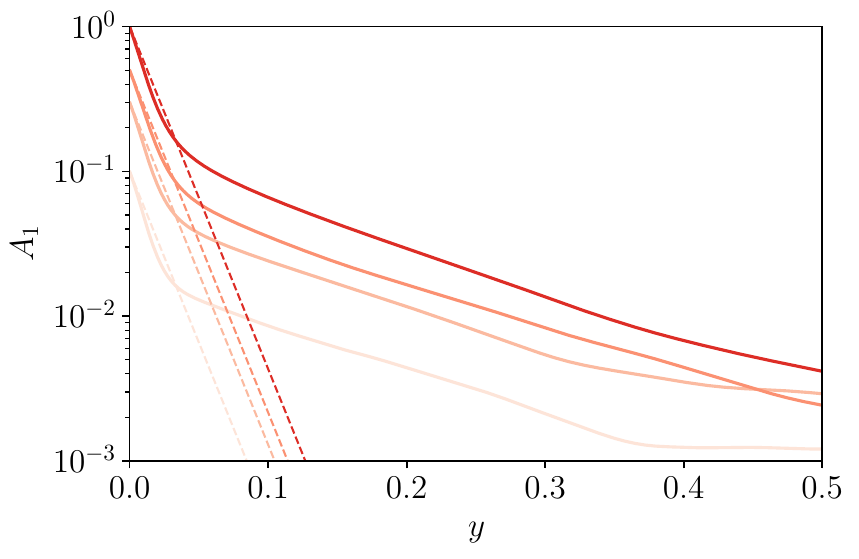}
  \includegraphics[width=0.45\textwidth]{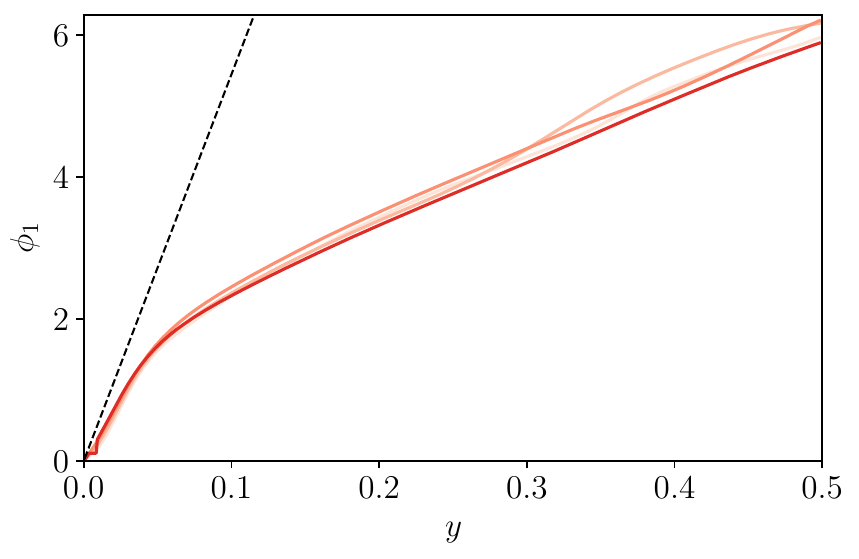}
  \caption{Amplitude (left) and phase (right) of modulation against wall distance for varying $\alpha$ and $T_\tau=0.8$. Lines are colored from smallest ($\alpha=0.1$) to largest ($\alpha=1$) amplitudes. Solid dashed lines indicate the respective Stokes solution.}
  \label{fig:a_phi_wo77}
\end{figure}

To confirm that the system's response to modulation can be understood through superposition, we check whether the response of the system is scaled if the modulation is scaled. To assess this, we conduct a series of simulations where we modulate the plate velocity with sinusoids of period $T_\tau=0.8$ and increasing amplitude ranging from $\alpha=0.1$ to $\alpha=1$. Figure \ref{fig:a_phi_wo77} presents the amplitude and phase response $A_1(y)$ and $\phi_1(y)$ from these runs. Notably, we find that the phase response is practically identical across all runs, while the amplitude response is undergoing a vertical shift- indicating a scaling of the response in the logarithmic scale used.

To finish, we examine the energy dissipation $\epsilon$. This quantity, usually examined in studies of homogeneous isotropic turbulence, takes even more relevance in shear flow as the time-averaged dissipation is directly proportional to the time-averaged frictional force at the plates \cite{eckhardt2007fluxes}. Increased dissipation is often observed in turbulent systems when the forcing is modulated close to the characteristic frequency of the flow. In thermal convection, methods have been proposed to enhance heat transport using this property, employing pulse-like perturbations \cite{jin2008experimental} or large sinusoidal perturbations \cite{yang2020periodically}. In the context of shear flow, this would mean increased transport of momentum. 

However, our simulations indicate that varying the forcing of the system has minimal effect. As mentioned earlier, across all studied waveforms at $\alpha=0.1$ the change in dissipation remains within the error bars of the unmodulated case. This also means that the time-averaged shear forcing at the plates remains unchanged despite the modulation, and shear transport cannot be enhanced this way.

\begin{figure}
 \includegraphics[width=0.45\textwidth]{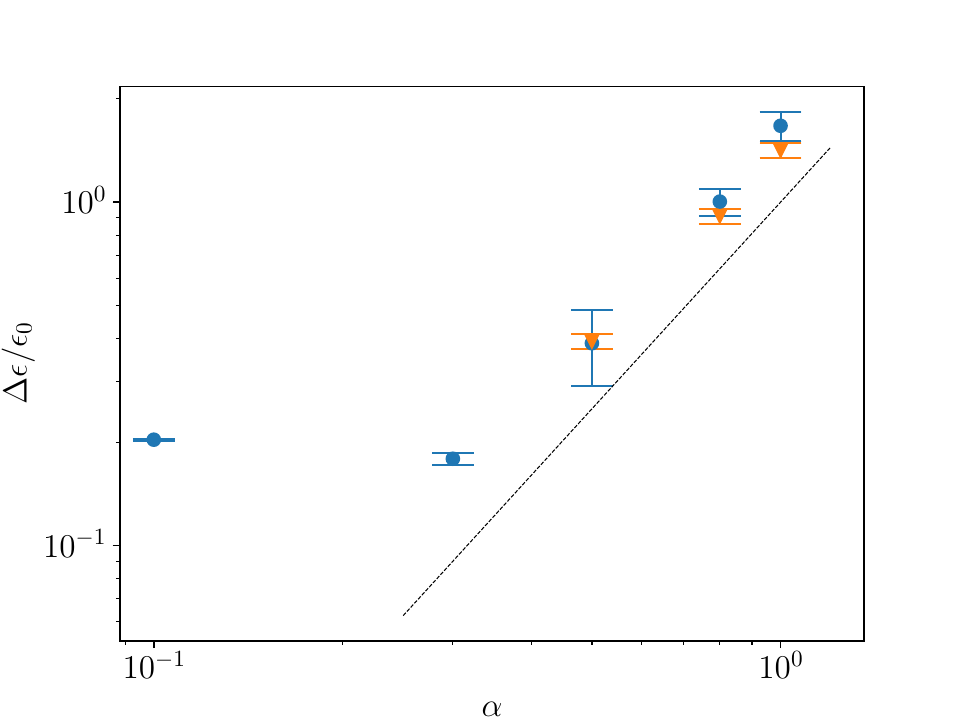}
  \caption{Quasilinear response in dissipation: Measured increase in dissipation $\Delta\epsilon/\epsilon_0$ against forcing amplitude $\alpha$ (blue circles), and increase predicted from simply adding the dissipation of a modulated case simulated separately (orange triangles). The black line marks a scaling of $\Delta\epsilon/\epsilon_0\sim\alpha^2$.}
  \label{fig:diss}
\end{figure}

Turning to the effects of forcing amplitude, Figure \ref{fig:diss} shows the relative dissipation change $\Delta \epsilon/\epsilon_0$ against $\alpha$, where $\Delta\epsilon=\epsilon-\epsilon_0$ and $\epsilon_0$ denotes the dissipation in the unmodulated scenario. For small values of $\alpha<0.3$, no significant increase in dissipation is observed. As the perturbation amplitude increases, the dissipation from the modulated flow becomes noticeable. However, this increase simply manifests as additional dissipation that can be computed separately and added to the base dissipation. The dissipation of the total flow can thus be estimated by summing the dissipation of the two constituent flows— one with constant forcing and the other with modulated forcing. Only when $\alpha=1$ (signifying modulation equal in magnitude to the constant forcing) does the dissipation of the combined flow marginally surpass the combined dissipation of the two base flows deviating from a summation correction.

Indeed, the approximate decomposition into an unaffected mean $\bar{U}_{x}$ and a modulated response remains valid until $\alpha$ approaches unity. Upon reaching sufficiently large forcing amplitudes, deviations from simple linear superposition become notable. However, at this point these are not unexpected. The system undergoes a change in dynamic response, as the modulated plate velocity overcomes the constant velocity, and may invert the local shear direction. Hence, while the occurrence of deviations is not surprising, the fact that they occur at relatively high amplitudes, far exceeding initial expectations, remains an intriguing result.

\section{Summary and Outlook}

To summarize, our study has demonstrated that shear flows exhibit responses to modulated perturbations which approximately satisfy the superposition principle under certain conditions. This finding is significant as it underscores that wall-bounded flows can manifest behaviors consistent with linear properties \cite{jimenez2013linear,mckeon2017engine}. To rationalize these findings, one might consider that a shear flow problem encompasses two distinct velocity scales, each generating different time-scales. The boundary layer, dominated by viscosity, operates under predominantly linear processes as the non-linear term is severely restricted. In high-Reynolds number wall-bounded flows, the relevant time-scales are notably slower than those of the turbulent bulk, and are typically on the order of $\mathcal{O}(d^2/\nu)$. In contrast, the bulk exhibits a continuum of time-scales, with large-scale structures having a characteristic time-scale of $\mathcal{O}(d/u_\tau)$. Because perturbations traversing the bulk must first transit the boundary layer, this generates a mismatch. If their time-scale is fast, matching the time-scales of the bulk, they are rapidly smoothed out. If their time-scale is slow, they are too slow to effectively couple with the time-scales existing in the bulk. The co-existence of two zones with distinct dynamics and time-scales complicates the potential for achieving resonances and greatly simplifies the dynamics of the system \cite{mckeon2017engine}. Indeed, this work is consistent with results which show that the large-scale modes of wall turbulence are dominant, and suggest that across a range of inputs the same dominant outputs will be produced \cite{farrell1993stochastic,bamieh2001energy,jovanovic2005componentwise}.

A remaining question is whether these results will translate to channel and pipe flows where the forcing comes from the bulk, rather than from the wall, and shear transport vanishes at the mid-gap (or the pipe axis). The modulation would be instantaneously felt by the entire flow, rather than through the wall where the non-linearity is much more constrained. Other possible extensions of this work are modulating the plates using spanwise velocities applied either on a single plate or on both. These types of modulation have been shown to substantially affect the properties of PCF \cite{bengana2022exact}, and could lead to unforseen interactions.

\emph{Acknowledgments:} We thank M. Fosas de Pando for helpful comments on the article.

\bibliography{apssamp}% Produces the bibliography via BibTeX.

\appendix 

\section{Box-size dependence studies}

To assess the adequateness of the computational box, an additional simulation with $L_x/d=4\pi$ and $L_z/d=2\pi$ and the 1-2-3 modulation was conducted. The number of points was doubled in the $x$ and $z$ direction to maintain the same resolution, but otherwise all other parameters were kept the same. This simulation was chosen as it was the one where the superposition of responses was most evident. 

\begin{figure}
 \includegraphics[width=0.45\textwidth]{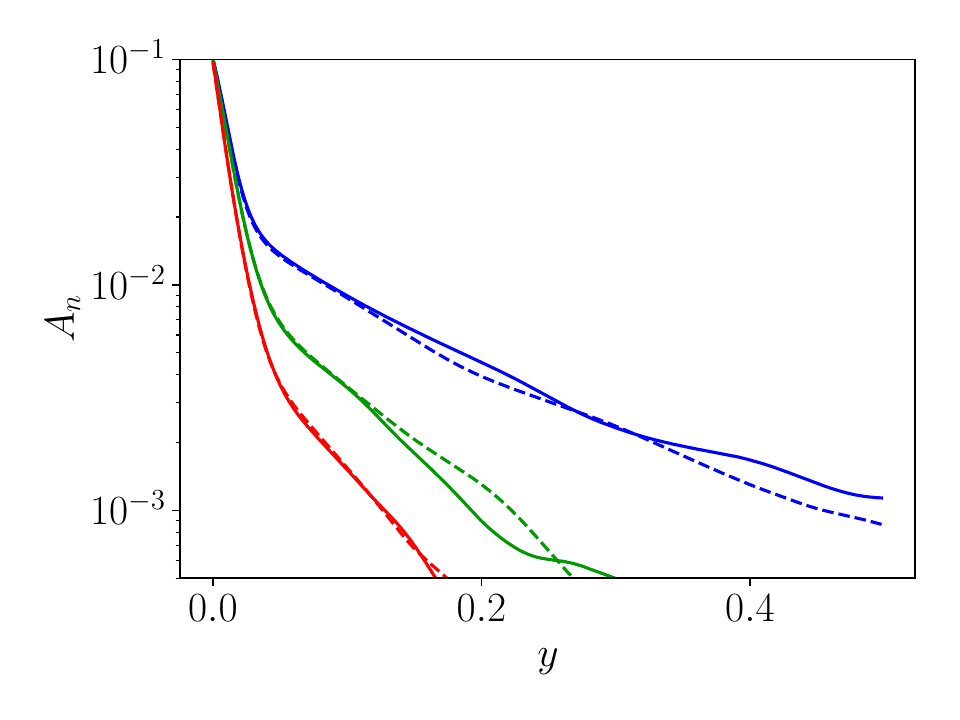}
 \includegraphics[width=0.45\textwidth]{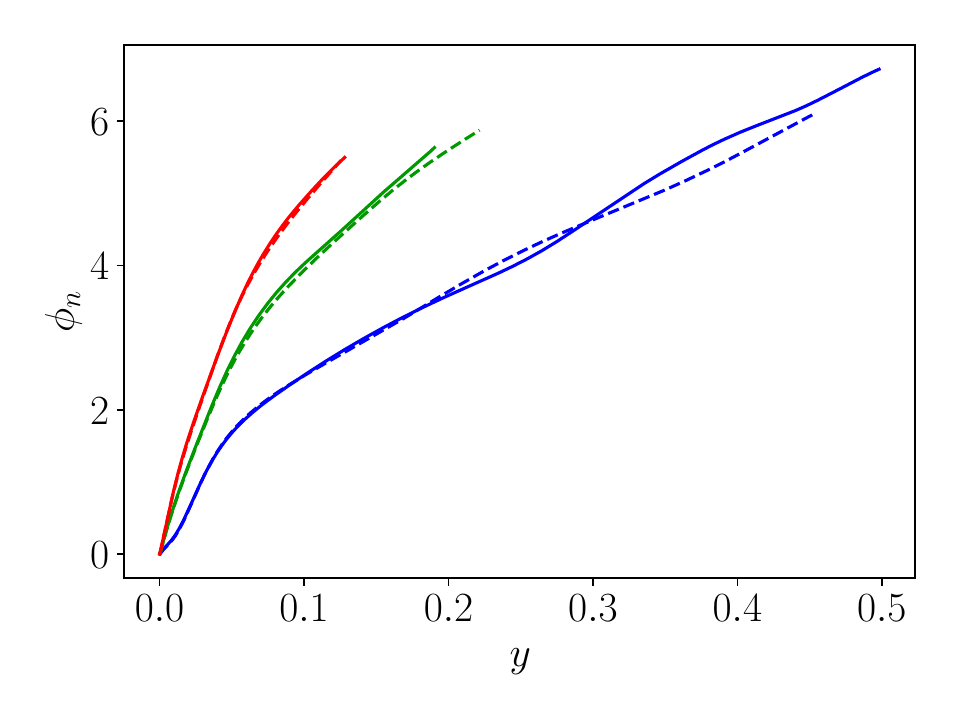}
\caption{Amplitude (left) and phase (right) response of the first three modes for the 1-2-3 simulation with $L_x/d=2\pi$ and $L_z/d=\pi$ (solid lines) and $L_x/d=4\pi$ and $L_z/d=2\pi$ (dashed lines). Symbols: blue, fundamental model; green, second harmonic; red, third harmonic.}
\label{fig:app-boxsize}
\end{figure}

Figure \ref{fig:app-boxsize} shows $A_n$ and $\phi_n$ for the first three harmonics for both the box size used in the manuscript, and the enlarged box size. It can be seen that aside from small deviations, especially when the amplitude becomes smaller, the basic physics of the problem is maintained. This shows that the computational box size used in this manuscript is enough to capture the physics of interest in this manuscript, i.e.~the superposition of responses. Additional information on box- or structure-size dependence of other quantities such as the dissipation and the energy spectra to a modulation in the plates spectra can be found in Ref.~\cite{akhtar2022effect}.

\end{document}